\documentstyle[12pt]{article}

\begin{document}

\begin{titlepage}
\begin{flushright}
{\large \bf CINVESTAV-FIS/97-14}
\end{flushright}
\vskip1.5cm
\begin{center}

{\Large \bf FCNC in leptonic and semileptonic decays of $D$  mesons
in a general two-Higgs doublet model}
 \vskip 2cm

{\large G. L\'opez Castro$^a$, R. Mart\'\i nez$^b$ and J. H.
Mu\~noz$^{a,c}$} \\

$^a$ {\it Departamento de F\'\i sica, Centro de Investigaci\'on y de  
Estudios} \\ {\it Avanzados del IPN, Apdo. Postal 14-740, 07000 M\'exico, 
D.F., M\'exico}

$^b$ {\it Departamento de F\'\i sica, A. A. 14490, Universidad}
\\ {\it Nacional, Bogot\'a, Colombia}

$^c$ {\it Departamento de F\'\i sica, Universidad del Tolima,} \\ 
{\it A.A . 546, Ibagu\'e, Colombia}
\end{center}
\vskip1.5cm
\begin{abstract}
Large long-distance standard model effects in  flavor-changing neutral
current (FCNC) semileptonic $D$ decays can make observable these processes
in future measurements. Eventual disagreements in this sector and/or
the observation of lepton family violating (LFV) $D$ decays would require
an explanation beyond the standard model  framework. In this paper we
confront present 
experimental data on leptonic and semileptonic FCNC and LFV $D$ 
meson decays with a version of the two-Higgs doublet model that 
allows these effects to occur at tree-level. The stringent 
bounds on the parameters of the model are obtained from $D^0 
\rightarrow l^+l'^-$ and $D \rightarrow \pi l^+l'^-$ decays. The 
consistency of the model requires that the branching fractions
of $D \rightarrow V l^+l'^-$ decays should be below the $10^{-9}$ level.
 \end{abstract}


\end{titlepage}%

\medskip

\

\

{\bf 1. Introduction}

  Flavor changing neutral currents (FCNC) in leptonic and semileptonic 
decays of charmed mesons are, higher order, very suppressed modes in the
standard model (SM) of particle interactions \cite{babu}--\cite{singer}.
The short-distance contributions to these processes in the SM are expected
to give branching fractions at the $10^{-19}$ level for $D^0 \rightarrow
\mu^+\mu^-$ and $10^{-9}$ for $D \rightarrow \pi l^+l^-$ processes, while
long-distance effects can enhance these predictions up to $10^{-15}$
\cite{pakvasa}  and $10^{-7} \sim 10^{-6}$ \cite{singer}, respectively.
Present experimental upper limits for these decays are in the range
$10^{-6}\sim 10^{-5}$ for 
$D^0 \rightarrow l^+l'^-$ and $10^{-3} \sim 10^{-5}$ for $D \rightarrow X 
l^+l'^-$ \cite{pdg}--\cite{cleo} ($X$ is a pseudoscalar or vector meson
and $l,\ l' = e,\ \mu$).
On the other hand, lepton family violating (LFV) processes, {\it i.e.} $l
\not = l'$,  are completely
forbidden in the SM scenario with unmixed lepton generations. 
Thus, FCNC and/or LFV leptonic
and semileptonic $D$ decays can serve to test the mechanisms responsible 
of long-distance contributions or eventually would require an explanation 
beyond the SM framework. Yet another (unlikely) possibility is that nature
places FCNC processes well below the SM expectations. This would force to
revise the estimates of long-distance effects or, again, to invoke
beyond the SM contributions to explain the eventual destructive
interference with the SM amplitudes.

   Recently, the study of FCNC in charm quark decays has attracted a
renewed  interest \cite{schwartz}--\cite{kaplan}. On the one hand, it
has been pointed out that these rare decays in models of new physics can
be enhanced over the SM predictions by several orders of magnitude
\cite{pakvasa}. On the other hand, the existing bounds on FCNC and LFV $D$ 
decays have been improved recently  at FERMILAB E791,
E771 and E687 experiments \cite{E791}--\cite{E687} and by the CLEO
Collaboration\cite{cleo}.
  In addition, some projects have been proposed with
the aim to reconstruct of the order
of $10^9$ charm decays during the Tevatron Run II \cite{kaplan} , which
would increase the sensitivity to FCNC and LFV processes by almost three
orders of magnitude with respect to present experiments. Therefore, it
becomes timely to explore all possible scenarios of new physics that may
give sizable contributions to these rare decays.

  In this paper we consider the constraints imposed by FCNC and LFV $D$
meson decays on a general two-Higgs doublet model that allows these
effects to contribute at tree level \cite{wolfenstein}. The variant of
the model considered here is built in such a way that
tree-level FCNC interactions of the neutral Higgses do not spoil the good
agreement between the SM predictions and experiment for the down quark
sector. The constraints on Yukawa interactions of the charged Higgses of 
this model have  
been studied in previous works \cite{nos}. Here we consider the effects of
Yukawa interactions of the neutral Higgses in FCNC and LFV decays of
$D$ mesons. To be more specific, we study the effects of neutral Higgses
of this model in the $D^0 \rightarrow l^+l'^{-},\ D \rightarrow P l^+l'^-$
and $D
\rightarrow Vl^+l'^-$ decays ($P(V)$ stands for a pseudoscalar (vector)
light meson and $l,\, l'=e$ or $\mu$), which will provide a rather wide
set of constraints on the effective Yukawa couplings of the model.

\

{\bf 2. The model.}

  The variant of the two-Higgs doublet model needed in our work has been
described elsewhere \cite{nos}. The general form of the Yukawa
interactions that allows tree-level FCNC processes is given by
\cite{wolfenstein} 
\begin{eqnarray}
{\cal L}_Y &=& 
\overline{Q_L^0}(F\tilde{\Phi}_1 + \xi F'\tilde{\Phi}_2)
U_R^0 + 
\overline{Q_L^0}(G\Phi_2 + \xi G'\Phi_1)D_R^0 \nonumber \\
& & + \overline{\Psi_L^0}(K\Phi_2 + \xi K'\Phi_1)l_R^0 + {\rm h.c.}, 
\end{eqnarray}
where $ F,\ F',\ G, \ G', K$ and $K'$ are dimensionless 3$\times$3
matrices, $\overline{Q_L^0} = (\overline{U^0_L}, \overline{D^0_L})$ with
$U^0_L$ ($D^0_L$) the triplet of left-handed up (down) quarks, and
$\overline{\Psi_L^0} = (\overline{\nu_L^0}, \overline{l_L^0})$ has a
similar definition in terms of leptonic fields. $\xi$
parametrizes the small breaking of the discrete symmetry that forbids
FCNC at tree-level. The superscript 0 in fermion fields stands for weak
eigenstates. 

 Since we are interested in having FCNC contributions only in the up-quark
sector, we shall drop the term proportional to $G'$ in Eq.
(1) \cite{nos}. Notice that the
Yukawa interactions for leptons are built to allow FCNC in the charged
leptons and keep massless neutrinos. After spontaneous symmetry breaking,
with $\langle \Phi_1 \rangle^T = (0,v_1/\sqrt{2})$ and $ \langle \Phi_2
\rangle^T = (0,v_2e^{-i\alpha'}/\sqrt{2})$, the model contains five
physical Higgses; the mass matrices for 
quarks and charged leptons become:
\begin{eqnarray}
M_U &=& \frac{1}{\sqrt {2}} (F v_1 + \xi F'v_2e^{-i\alpha'})\ , \\
M_D &=& \frac{1}{\sqrt {2}} G v_2\  , \\
M_l &=& \frac{1}{\sqrt {2}} (K v_2 + \xi K' v_1e^{-i\alpha'})\ .
\end{eqnarray}

  For simplicity we choose to work in a basis where $M_U$ and $M_l$ are
diagonal. Notice that, unlike the case where $\xi=0$, $F$ and $F'$
(respectively, $K$ and $K'$) are not diagonal matrices and can allow for
(unsuppressed by fermion masses) FCNC interactions in the up-quark sector.
 
 The Yukawa interactions between mass eigenstates of  neutral
scalar Higgses ($H_0$ and $h_0$), the pseudoscalar Higgs ($A_0$)  and
 the fermions ($U=(u,\ c,\ t)$ and $l=(e, \mu, \tau)$) are given by (we do
not write the interactions of down
quarks because we are interested in FCNC in the up sector): 
\begin{eqnarray}
{\cal L}_N &=& 
\frac{1}{\sqrt{2}} \overline{U}\left\{ (F\cos \alpha + \xi F' \sin
\alpha) H_0 
+ (-F \sin \alpha + \xi F' \cos \alpha) h_0 \right. \nonumber \\
& & \ \left. + i(F\sin \beta -\xi F'\cos\beta) A_0 \gamma_5 \right \} U
\nonumber \\
& & \ + \frac{1}{\sqrt{2}} \overline{l}\left\{ (K\sin \alpha + \xi K' \cos
\alpha) H_0 
+ (K \cos \alpha - \xi K' \sin \alpha) h_0 \right. \nonumber \\
& & \ \left. + i(K\cos \beta -\xi K'\sin\beta) A_0 \gamma_5 \right \} l\ .
\end{eqnarray}

  In these expressions, $\alpha$ is the angle that appears in the
diagonalization of the neutral scalar Higgs bosons and $\tan \beta\equiv
v_2/v_1$ .

  Due to the low energy scales involved in charm meson decays it becomes
convenient to write out an effective four-fermion interaction Hamiltonian 
to describe the tree-level processes of our interest. The form of this
Hamiltonian is:
\begin{equation}
{\cal H}_{eff} =\frac{G_F}{\sqrt{2}} \left\{ 
\overline{U} \Lambda_{H_0} U \cdot \overline{l} L_{H_0} l + 
\overline{U} \Lambda_{h_0} U \cdot \overline{l} L_{h_0} l +   
\overline{U} \Lambda_{A_0} \gamma^5 U \cdot \overline{l} L_{A_0} \gamma_5l
\right\} \ .
\end{equation}

 Using Eqs. (2) and (4), the effective couplings $\Lambda_i$ and $L_i$
can be written as:
\begin{eqnarray}
\Lambda_{H_0} &=& \frac{2m_W}{gm_{H_0}} \xi F' (\sin \alpha - \tan\beta
\cos\alpha e^{-i\alpha'} )\ , \\
\Lambda_{h_0} &=& \frac{2m_W}{gm_{h_0}} \xi F' (\cos \alpha + \tan\beta
\sin \alpha e^{-i\alpha'} )\ , \\ 
\Lambda_{A_0} &=& -\frac{2m_W}{gm_{A_0}} \xi F' (\cos \beta + \tan\beta
\sin \beta e^{-i\alpha'} ) \ , \\ 
L_{H_0} &=& \frac{\sqrt{2} M_l}{m_{H_0}}
\frac{\sin\alpha}{\sin \beta} + \frac{2m_W}{gm_{H_0}} \xi K' (\cos \alpha
- \cot\beta \sin\alpha e^{-i\alpha'} )\ , \\
L_{h_0} &=& \frac{\sqrt{2} M_l}{m_{h_0}}
\frac{\cos\alpha}{\sin \beta} - \frac{2m_W}{gm_{h_0}} \xi K' (\sin \alpha
+ \cot\beta \cos\alpha e^{-i\alpha'} )\ , \\
L_{A_0} &=& \frac{\sqrt{2} M_l}{m_{A_0}}\cot \beta 
- \frac{2m_W}{gm_{A_0}} \xi K' (\sin \beta
+ \cot\beta \cos\beta e^{-i\alpha'} )\ .
\end{eqnarray}

As already anticipated, the leptonic couplings contain a (diagonal) piece
proportional to fermion masses\footnote{ Since we are interested in $ c 
\rightarrow u$ transitions we do not write a corresponding diagonal
mass term in the quark couplings.} and another (non-diagonal) piece 
which is not {\it a priori} suppressed by fermion masses and will induce
FCNC interactions. 

  If we assume a specific {\it ansatz} for the Yukawa couplings $F'$ and
$K'$, we can use the experimental data on $D$ decays to get bounds on the
remaining parameters of the model. Instead, in the following we choose to
use the available data to constrain the effective couplings given in Eqs.
(6)--(12).

\

{\bf 3. Constraints from leptonic and semileptonic $D$ decays.}

  The relevant hadronic matrix elements of the $\overline{u}c$ and
$\overline{u}\gamma_5 c$ currents can be computed from the
divergence of the $ c \rightarrow d$ vector and axial vector
charged currents  
and using isospin symmetry. Thus, we obtain:
\begin{eqnarray}
\langle 0 | \overline{u}\gamma_5 c| D^0(p) \rangle &=& if_D
 \frac{m_D^2}{m_c+m_u}\ , \\
\langle \pi^+(p') | \overline{u} c| D^+(p) \rangle &=& 
\sqrt{2}\, \langle \pi^0(p') | \overline{u} c| D^0(p) \rangle \ ,  \\ &=&
\left (\frac{m_D^2-m_{\pi}^2}{m_c-m_u}\right) F^{D^0 \rightarrow
\pi^-}_0(q^2)\ , \\
\langle V(p',\varepsilon^*) | \overline{u} c| D (p)\rangle &=& 0 \ , \\
\langle \rho^+(p',\varepsilon^*) | \overline{u}\gamma_5 c| D^+(p) \rangle
&=&
\sqrt{2}\, \langle \rho^0 (p',\varepsilon^*)| \overline{u}\gamma_5 c| D^0
(p) \rangle \ ,\\
&=&-\, \frac{2im_{\rho}}{m_c +m_u} q.\varepsilon^* A^{D^0 \rightarrow
\pi^-}_0(q^2)\ ,
\end{eqnarray}
where $q=p-p'$ is the momentum transfer to the lepton pair and
$\varepsilon^*$ is the polarization four-vector of the outgoing vector
meson. In Eq. (16) $V$ is a vector meson. Notice that the matrix elements
for the $D \rightarrow P$ and $D\rightarrow V$ transitions depend on only
one form factor at the time. This happens because only the relative wave
$l=0$ and $l=1$ of the $P$-Higgs and $V$-Higgs systems contribute to
these transitions, respectively.

  For the $D$ meson decay constant we take the value $f_D =217$ MeV
which is obtained from the relation $f_D/f_{D_s} \approx 0.9$ \cite{bls}
and $f_{D_s} = 241$ MeV from \cite{reff_D}. The $q^2$-dependence of the
scalar and pseudoscalar
form factors appearing in Eqs. (14)--(18) are chosen to be  monopolar
\begin{equation}
F_0(q^2) = \frac{F_0(0)}{1-q^2/m_{0^+}^2}\ , \ \ \ \ \ \ 
A_0(q^2) = \frac{A_0(0)}{1-q^2/m_{0^-}^2}\ , 
\end{equation}
where $m_{0^+}$ and $m_{0^-}$ are the masses of the scalar and
pseudoscalar neutral $D$ mesons, respectively. The normalizations of these
form factors at $q^2=0$ are taken from the relativistic quark model of
 Wirbel-Stech-Bauer \cite{wsb}.

   The other hadronic matrix elements needed in our calculation are fixed
 either by identifying the $\overline{u}u$ content of final state
isosinglet mesons, namely:
\begin{eqnarray}
\langle \eta | \overline{u} c| D^0 \rangle &=& 
\frac{1}{\sqrt{3}}\langle \pi^0 | \overline{u} c| D^0
\rangle\,(\cos\theta_P-\sqrt{2}\sin \theta_P) \ , \\ 
\langle \omega | \overline{u} \gamma_5c| D^0 \rangle &=& 
\langle \rho^0 | \overline{u}\gamma_5 c| D^0 \rangle \ , 
\end{eqnarray}
or using SU(3) flavor symmetry:
\begin{eqnarray}
\langle K^+ | \overline{u} c| D^+_s \rangle &=& 
\langle \pi^+ | \overline{u} c| D^+ \rangle \ , \\ 
\langle K^{*+} | \overline{u} \gamma_5c| D^+_s \rangle &=& 
\langle \rho^+ | \overline{u}\gamma_5 c| D^+ \rangle \ . 
\end{eqnarray}
Notice that we assume ideal $\omega-\phi$ mixing and we use $\theta_P =
-20^0$ in Eqs. (20)--(21).

The information on the experimental data about the FCNC and LFV $D$ decays
is taken from the 1997 update of Ref. \cite{pdg}, which already
incorporates some recent results of Refs. \cite{E791}--\cite{cleo}. 

   In Table 1 we show the upper bounds for the products of couplings
constants that can be constrained from the experimental data considered.
 We have introduced in Table 1 a short notation for coupling
constants. First, we express the bounds from leptonic $D^0$ decays and $D
\rightarrow V l^+l'^-$ decays in
 terms of $\alpha^{ll'}\equiv \Lambda^{uc}_{A_0} L^{ll'}_{A_0}$. 
 Since both scalar neutral Higgses contribute to $D \rightarrow P l^+l'^-$
we have expressed the upper bounds in terms of the quantity $\sigma^{ll'}
\equiv \Lambda^{uc}_{H_0} L^{ll'}_{H_0} +\Lambda^{uc}_{h_0}
L^{ll'}_{h_0}$. 
 
 Despite the fact that all the upper limits on branching ratios are at the
$10^{-4}\sim 10^{-5}$ level, the different bounds on the effective
couplings spread over two orders of magnitude. 
  From Table 1 we conclude that the stronger bounds on the $\alpha^{ll'}$
couplings come from purely leptonic $D^0$ decays, while the same
 bounds from $D \rightarrow V l^+l'^-$ decays are rather weak.
Therefore, in the context of the present model, the leptonic $D^0$
decays imply that branching ratios of three-body decays of $D$'s involving
vector mesons should be below the $10^{-9}$ level. On
the other hand, the best constraints on the $\sigma^{ll'}$ couplings are
obtained from the $D \rightarrow \pi l^+l'^-$ mainly because of the phase
space suppression in the decays involving the $\eta$ meson. Finally, since
the $V$-Higgs system in $D \rightarrow Vl^+l'^-$ decays is in a $l=1$
relative wave, this gives a further phase space suppression and the
absolute numerical bounds on the $\alpha^{ll'}$'s becomes weaker than the
limits on the $\sigma^{ll'}$'s (obtained from $D \rightarrow P$
transitions).

  In order to draw any information on the Yukawa couplings of our interest
let us make some considerations. To start, let us neglect the first
term\footnote{Notice that this approximation is not necessary in the case
of LFV decays.} in Eqs. (10)-(12) and set $\alpha'=0$. In this case we
obtain the following expressions for $\alpha^{ll'}$ and $\sigma^{ll'}$:
\begin{eqnarray}
\alpha^{ll'}&=&\frac{1}{\sqrt{2}G_Fm_{A_0}^2} 
\frac{(\xi F')^{uc} (\xi K')^{ll'}}{\sin \beta \cos \beta} \ , \\
\sigma^{ll'}&=& -\, \frac{1}{\sqrt{2}G_F} 
\frac{(\xi F')^{uc} (\xi K')^{ll'}}{\sin \beta \cos \beta} \left \{
\frac{\sin^2(\alpha -\beta)}{m_{H_0}^2}+
\frac{\cos^2(\alpha-\beta)}{m_{h_0}^2} \right\}\ , 
\end{eqnarray}
or the relationship
\begin{equation}
\sigma^{ll'} \leq -\, m_{A_0}^2 
\left \{ \frac{1}{m_{H_0}^2}+\frac{1}{m_{h_0}^2} \right\}\alpha^{ll'} \ .
\end{equation}

  In the absence of information regarding the parameters of
this model we will assume $\tan \beta \approx 1,\ m_{h_0} = 130\ {\rm
GeV}$
and $m_{H_0}=m_{A_0} = 300$ GeV. From Eq. (24) and the bounds on
$\alpha^{ll'}$
obtained from leptonic $D^0$ decays (see Table 1)  we derive:
\begin{eqnarray}
(\xi F')^{uc} (\xi K')^{ee} &\leq & 2.9 \times 10^{-3} \\
(\xi F')^{uc} (\xi K')^{\mu \mu} &\leq & 1.7 \times 10^{-3} \\
(\xi F')^{uc} (\xi K')^{\mu e} &\leq & 3.6 \times 10^{-3}.
\end{eqnarray}
Therefore, one may conclude that present experimental data on FCNC and LFV
$D$ decays only mildly constrain the strength of products of the
relevant Yukawa couplings of this model. Since the (diagonal) terms
proportional to fermion masses in Eqs. (10)--(12) are of $O(10^{-4})$ for
the $D \rightarrow X \mu^+ \mu^-$ modes, the approximation done to derive
Eqs. (24)--(25) is justified in view of the present experimental upper
limits.

Note that if a specific ansatz is assumed for these Yukawa couplings
\cite{ansatz}, then Eq. (24) can furnish the allowed region for $m_{A_0}$
as a function of $\beta$.  Let us notice however, that Eq. (25) does not
provide additional constraints on the Yukawa couplings unless, in
addition, some information on the mixing angle $\alpha$ is introduced by
hand. 

  In summary, in this work we have studied the constraints imposed by FCNC
and LFV leptonic and semileptonic $D$ decays on a version of the two-Higgs
doublet model that contains these effects at tree-level. The stringent
bounds on the relevant Yukawa couplings are obtained from two-body
leptonic $D^0$ decays which are mediated by the pseudoscalar Higgs boson
of the model. The best constraints on the Higgs scalar interactions are
obtained from $D \rightarrow \pi l^+l'^-$ decays. The three-body $D$ 
decays involving vector mesons provide only very weak bounds and their
measurements would have to be improved by five orders of magnitude in
order to furnish similar constraints on the model as obtained from purely  
leptonic decays.

\

{\bf Acknowledgements}. The authors are grateful to Conacyt (G.L.C.) and
Colciencias (R.M. and J.H.M.) for financial support.

\newpage

\begin{center}
\begin{tabular}{|c|c|c|}
 \hline
Channel& Exp. BR & upper bound \\ 
\hline
$D^0 \rightarrow e^+e^-$ & $<1.3 \times 10^{-5}$ & $\alpha^{ee}<4.0 \times
10^{-3}$ \\
$D^0 \rightarrow \mu^+\mu^-$ & $< 4.2  \times 10^{-6}$  & $\alpha^{\mu\mu}
<2.3 \times 10^{-3}$   \\
$D^0 \rightarrow \mu^{\pm}e^{\mp}$ & $<1.9 \times 10^{-5}$ & $\alpha^{\mu
e} <4.9 \times 10^{-3}$  \\
\hline
$D^0 \rightarrow \pi^0 e^+e^-$ & $<4.5 \times 10^{-5}$ & $\sigma^{ee}<4.2 
\times 10^{-2}$ 
\\
$D^0 \rightarrow \pi^0 \mu^+\mu^-$ & $<1.8 \times 10^{-4}$ & $\sigma^{\mu
\mu} <8.6 \times 10^{-2}$ \\
$D^0 \rightarrow \pi^0 \mu^{\pm}e^{\mp}$ & $<8.6 \times 10^{-5}$ &
$\sigma^{\mu e}<5.8 \times 10^{-2}$ \\
$D^0 \rightarrow \eta e^+e^-$ & $<1.1 \times 10^{-4}$ & $\sigma^{ee}<0.16$
\\
$D^0 \rightarrow \eta \mu^+\mu^-$ & $<5.3 \times 10^{-4}$ &
$\sigma^{\mu\mu}<0.38$ \\
$D^0 \rightarrow \eta \mu^{\pm}e^{\mp}$ & $<1.0 \times 10^{-4}$ &
$\sigma^{\mu e}<0.16$ \\
$D^+ \rightarrow \pi^+ e^+e^-$ & $<6.6 \times 10^{-5}$ & $\sigma^{ee}<2.2 
\times 10^{-2}$  \\
$D^+ \rightarrow \pi^+ \mu^+\mu^-$ & $<1.8 \times 10^{-5}$ &
$\sigma^{\mu\mu}<1.2 \times 10^{-2}$ \\
$D^+ \rightarrow \pi^+ \mu^-e^+$ & $<1.1 \times 10^{-4}$ & $\sigma^{\mu
e}<2.9 \times 10^{-2}$ \\
$D_s^+ \rightarrow K^+ \mu^+\mu^-$ & $<5.9 \times 10^{-4}$ &
$\sigma^{\mu\mu}<0.15$ \\
\hline
$D^0 \rightarrow \rho^0 e^+e^-$ & $<1.0 \times 10^{-4}$ &
$\alpha^{ee}<0.35$
\\
$D^0 \rightarrow \rho^0 \mu^+\mu^-$ & $<2.3 \times 10^{-4}$ &
$\alpha^{\mu\mu}< 0.57$ \\
$D^0 \rightarrow \rho^0 \mu^{\pm}e^{\mp}$ & $<4.9 \times 10^{-5}$ &
$\alpha^{\mu e}<0.25$ \\
$D^0 \rightarrow \omega e^+e^-$ & $<1.8 \times 10^{-4}$ &
$\alpha^{ee}<0.48$
\\
$D^0 \rightarrow \omega \mu^+\mu^-$ & $<8.3 \times 10^{-4}$ &
$\alpha^{\mu \mu}<1.14$ \\
$D^0 \rightarrow \omega \mu^{\pm}e^{\mp}$ & $<1.2 \times 10^{-4}$ &
$\alpha^{\mu e}<0.40$ \\
$D^+ \rightarrow \rho^+ \mu^+\mu^-$ & $<5.6 \times 10^{-4}$ &
$\alpha^{\mu\mu}<0.39$ \\
$D_s^+ \rightarrow K^{*+} \mu^+\mu^-$ & $<1.4 \times 10^{-3}$ &
$\alpha^{\mu\mu}<0.96 $ \\
\hline
\end{tabular}
\end{center}
\begin{center}
Table 1. Bounds on Yukawa couplings from FCNC and LFV $D$ meson decays.
\end{center}

\end{document}